\documentclass[aps,pra,twocolumn,superscriptaddress,floatfix]{revtex4-2}

\usepackage[utf8]{inputenc}
\usepackage[english]{babel}
\usepackage{mathtools}
\usepackage{physics}
\usepackage{float}
\usepackage{braket}
\usepackage{xcolor}
\usepackage{bbm}
\usepackage{bm}
\usepackage{amsbsy}
\usepackage{amsthm}
\usepackage{amssymb}
\usepackage{amsfonts}
\usepackage{amsmath}
\usepackage{dsfont} 
\usepackage{graphicx} 
\usepackage{hhline}
\usepackage{epsfig}
\usepackage{epstopdf}
\usepackage{dsfont}
\usepackage{multibib}
\usepackage{xcolor}
\usepackage{color}
\usepackage{verbatim}
\usepackage{nomencl}
\usepackage[colorlinks]{hyperref}
\usepackage{float}
\usepackage[normalem]{ulem}
\usepackage{multirow}
\usepackage{makecell}

\def\>{\rangle}
\def\<{\langle}
\def\tr{{\rm Tr}}

\begin{document}

\title{Smooth Crossover Between Weak and Strong Thermalization using Rigorous Bounds on Equilibration of Isolated Systems}
\author{Luis Fernando dos Prazeres}
\affiliation{Instituto de Fíisica, Universidade Federal Fluminense, Av. Litoranea s/n, Gragoatá, 24210-346 Niterói, Rio de Janeiro, Brazil}
\author{Thiago R. de Oliveira}
\affiliation{Instituto de Fíisica, Universidade Federal Fluminense, Av. Litoranea s/n, Gragoatá, 24210-346 Niterói, Rio de Janeiro, Brazil}

\begin{abstract}

It is usually expected and observed that non-integrable isolated quantum systems thermalize. However, for some non-integrable spin chain models, in a numerical study, initial states with oscillations that persisted for some time were found and the phenomenon was named weak thermalization. Later, it was argued that such oscillations will eventually decay suggesting that weak thermalization was about time scales and not the size of the fluctuations. Nevertheless, the analyses of the size of the fluctuations were more qualitative. Here, using exact diagonalization we analyze how the size of the typical fluctuation, after long enough time for equilibration to happen, scales with the system size. For that, we use rigorous mathematical upper bounds on the equilibration of isolated quantum systems. We show that weak thermalization can be understood to be due to the small effective dimension of the initial state. Furthermore, we show that the fluctuations decay exponentially with the system size for both weak and strong thermalization indicating no sharp transitions between these two regimes.
\end{abstract}

\maketitle

\section{Introduction}

Although very successful in explaining a large range of phenomena, statistical mechanics still raises debates about its foundations. In particular, how to justify the use of ensembles to describe single systems and to show that macroscopic systems reach an equilibrium thermal state, even though the laws governing the motion of the microscopic constituents have time-reversal symmetry. In the classical scenario, the most known argument to solve some of these problems is chaos and ergodicity: to show that the system dynamics explores all, or almost all, the accessible phase space and thus time averages, that occur in real experiments, agree with ensemble averages used in calculations \cite{Haar95}. In the quantum case, there are also many works trying to establish such a connection. However this approach has its problems: for example, it is hard to prove that a system is ergodic, and the time scales to have ergodic behavior are much larger than usual physical time scales (see Sec. I of \cite{Borgonovi16} and \cite{Alessio16}).

Another approach, usually named Typicality, does not involve dynamics. One shows that most, or typical, microscopic configurations correspond to equilibrium macroscopic states. This approach was proposed already by Boltzmann \footnote{See \cite{Lebowitz93, Lebowitz07}
and Sec. 8 of \cite{Brown09} for a discussion about how Boltzmann changed his arguments to justify the second law} and extended to the quantum scenario by von Neumann, showing that most pure quantum states correspond to an equilibrium macroscopic state \cite{Goldstein10}. With Typicality, it is argued that one does not need to consider ensembles (mixed states) and it is possible to show equilibration even for individual isolated systems described by point in the phase space or pure states in the Hilbert space. There are still debates about the best approach (see \cite{Lebowitz93}, the comments, reply, and \cite{Singh13}). Nonetheless, in the last decade, many works revived this approach by studying Typicality from a quantum information perspective \cite{Popescu06, Gogolin16, Mori18}. Furthermore, if most states appear
at equilibrium, it is expected that generic dynamics, and not only ergodic ones, will lead to equilibration. It is possible to show that isolated quantum systems, described by a pure quantum state and evolving unitarily, in general, equilibrate. Most of these recent works deal with the problem from an abstract and formal perspective: proving general theorems about mathematical properties of generic Hamiltonian and initial states for equilibration to happen \cite{Reimann2008, Linden08, Gogolin16}. While these works are very general and rigorous, they do not give much information about the relevant physical properties of equilibration.

The last decade also saw a revival of the role of chaos in thermalization with the introduction of the Eigenstate Thermalization Hypothesis (ETH), where each single energy eigenstate corresponds to an equilibrium macroscopic state. With ETH, one can show that isolated quantum systems evolving unitarily will reach not only equilibrium but thermal equilibrium. 
While one can only prove ETH for random Hamiltonians, many numerical studies showed that it is usually true for non-integrable systems with hints of quantum chaos. Thus, many numerical works considering a quench on specific Hamiltonian corroborated the classical expectation that non-integrability and chaos will lead to thermal equilibrium; even for isolated systems \cite{Alessio16, Borgonovi16}. 
Although most of these analyses showed thermal equilibrium for non-integrable systems there are exceptions. In particular, in \cite{Banuls2011} it was shown that for a non-integrable $H$ some initial states do not equilibrate to a thermal state, but keep oscillating around the thermal state; this was named weak thermalization. Later, it was proposed that this weak thermalization could be understood in terms of quasi-particles excitations and that the weak thermalization was due to a small number of weak interacting quasi-particles \cite{Chengju2017}. It was also argued that eventually, these initial states would equilibrate, but at longer times. In \cite{Hyungwon15} it was shown that there are local operators whose fluctuations decay slowly and that this could explain the weak thermalization of \cite{Banuls2011}. Finally, in \cite{Farrelly17} it was shown that for local Hamiltonians initial states with finite energy variance and exponentially decaying correlations will equilibrate after a long enough time; thus equilibration should also occur in the weak thermalization scenario. 
Therefore, there is strong evidence that the phenomenon of weak thermalization is mainly about the time scales for the fluctuations to become small. However, to better characterize different regimes of equilibration and thermalization it is important to have a more quantitative analysis of the size of the fluctuations. One important aspect is how the size of typical fluctuations, decays with the system size, is there any particular behavior in the regime of weak thermalization? We should also mention that weak thermalization was also found in other models \cite{Sun21, Lin19} and even studied experimentally \cite{Fshen2021}.

It would thus be interesting to connect the complementary results of these two parallel lines of research. In the present work, we try to establish such a connection showing that weak thermalization can be easily understood in terms of the effective dimension that appears in general mathematical theorems about equilibration. We also improve the understanding of weak thermalization by studying, numerically with exact diagonalization and systems up to 15 spins, the scaling of the size of fluctuations with the system size: we show that in both cases the fluctuations decay exponentially but at different rates. Thus it seems there is a smooth crossover from weak to strong thermalization.

\section{Rigours Theorems about Equilibration}

In the more abstract approach, one considers a very general scenario of an isolated system, with Hamiltonian $H=\sum_n E_n |E_n\>\<E_n|$, in a given non-equilibrium initial state $|\psi_0 \>=\sum_n c_n |E_n\>$. As the system
is isolated, its state after a time $t$ is given by
\begin{equation}
\label{eq:evol}
    |\psi(t) \>= \mathcal{U}(t) |\psi_0\>,
\end{equation}
with
\begin{equation}
 \mathcal{U}(t) = e^{-iHt}.
 \end{equation}

The main question is if the system will equilibrate and if the equilibrium state is a thermal one. As the evolution is unitary, $|\psi(t)\>$ will never stop evolving and approach a steady equilibrium state.
One could look at some observable $A$ or a subsystem $\rho_S= \tr_B \left[ |\psi\>\<\psi| \right ]$. These quantities, will also never stop evolving for finite systems, but they may oscillate around an equilibrium value with undetectable fluctuation and therefore be at equilibrium for all practical purposes. To quantify the size of the fluctuation one considers the average fluctuation, given by
\begin{equation}
\overline{\Delta A^2} = \overline{(\<A(t)\> - \overline{\<A\>})^2},
\end{equation}
with the overline being a time average; for example
\begin{equation}
 \overline{\<A\>}  = \lim_{T\to \infty  }  \frac{1}{T} \int_0^T dt \;\; \langle \hat A(t) \rangle .
\end{equation}
In the case one looks at a sub-system, it is possible to study the average distance between $\rho_S$ and an equilibrium sub-system state $\overline{\rho}_S$. In both cases, if the average fluctuation is small the system is most of the time, or typically, close enough to an equilibrium value and for all practical purposes is indistinguishable from it.

For non-degenerated $H$ it is simple to show that
\begin{equation}
\overline{\rho} = \sum_n |c_n|^2 |E_n\>\<E_n|
\end{equation}
and therefore $\overline{\rho}_S=\tr_B \left[ \overline{\rho} \right] $ and 
$ \overline{\<A\>}= \tr \left[ A \overline{\rho} \right] $. The equilibrium, and
average, state is the dephased initial state and is also referred to as the diagonal ensemble. 

It is also possible to obtain an upper bound on the average fluctuation from the equilibrium state. It is possible to show that \cite{Reimann2008, Linden08}

\begin{equation}
\overline{(\<A(t)\> - \overline{\<A\>})^2} \leq ||A|| / d_{eff},
\end{equation}
with
\begin{equation}
d_{eff}=\frac{1}{\sum |c_n|^4} = \frac{1}{\tr(\overline{\rho}^2)},
\label{deff}
\end{equation}
the effective dimension. $d_{eff}$ is a metric used to quantify the number of energy eigenstates involved in the initial state $\ket{\psi_0}$  or the degree to which the corresponding Hilbert space is explored during the process of time evolution. It is worth noting that $1/d_{eff}$ sometimes is also referred to as the inverse participation ratio (IPR), which measures the degree of localization of a given state on an energy basis. This result was first obtained for Hamiltonians that lack degeneracies and "degenerate gaps.": $E_k-E_l=E_m-E_n$ is valid only if $E_k=E_m; E_l=E_n$ or $E_k=E_l; E_m=E_n$.  Although it is generally expected and numerically observed that such conditions hold for systems of interacting particles, the results may be extended to systems where the number of degeneracies and "degenerated gaps" is not exponentially large \cite{Short11} and also for infinite-dimensional systems \cite{Reimann12}.

These results give the mathematical conditions for an initial state $|\psi_0\rangle$, $H$, and $A$ to equilibrate but do not address the question of the nature of the equilibrium state: if it is thermal or not. Actually, in general, we do not expect the equilibrium state $\overline{\rho}$ to be thermal; the distribution of the $|c_n|^2$ is not of the micro or canonical form and contains information about the initial state. Thus we have to separate the question of equilibration of the systems, if the fluctuations are small, from the thermalization that also considers the form of the equilibrium state. Note that, it is possible to show that if one randomly chooses an initial state in the Hilbert space, most states will equilibrate to the same sub-system state \cite{Linden08}. One can also argue, that for local $H$ the distance between energy levels becomes exponentially small with the system size, and therefore most states prepared in the lab will have large $d_{eff}$ and equilibrate \cite{Reimann2008}. Last, it is possible to reason that when the equilibrium state is not the thermal one, it is still the more fundamental (\cite{Reimann2008}).

The equilibration can be simply understood as a dephasing. The time
evolution of the observable is given by 
\begin{equation}\label{eq:observable_time}
\langle \hat A (t) \rangle =  \overline{\langle A \rangle} + \sum_{n \neq m}^N  \langle E_n |A|E_m\rangle \  c_n^* c_m e^{-i(E_m - E_n)  t}.
\end{equation}
We can see that the time evolution is given by a sum of complex numbers with phases that change in time by a "velocity" $E_n-E_m$.
All these complex numbers (phasors) needed to be almost aligned for a non-equilibrium initial state. But as time evolves each complex number gets a different phase and, for incommensurable gaps, the phasors spread uniformly in the complex plane becoming an equilibrium state. It is possible to estimate time scales for the equilibration \cite{Tro2018, Short11, Wilming18}. However, as we have a quasi-periodic function revivals very close to the initial state will occur, although its time scale increases exponentially with the system size \cite{Peres82, Venuti15}.

Therefore we know that for $H$ with not many "degenerate gaps", initial states with large $d_{eff}$ will equilibrate when one looks at observable with a not large $||A||$. Note that the upper bound gives no information about the time scales, but only the guarantee that equilibration will happen after enough time.
Actually, one should study how $||A||$ and $d_{eff}$  scale with the system size $N$, since we only expect equilibration for large systems. One drawback is that while the results are rigorous and very general they do not give information about more physical properties of the system, initial state, and observables nor if the the equilibrium state is thermal;
with a few exceptions as \cite{Farrelly17}.

\section{Thermalization, Integrability, and Chaos}

\begin{figure*}[t]
\includegraphics[width=.45\textwidth]{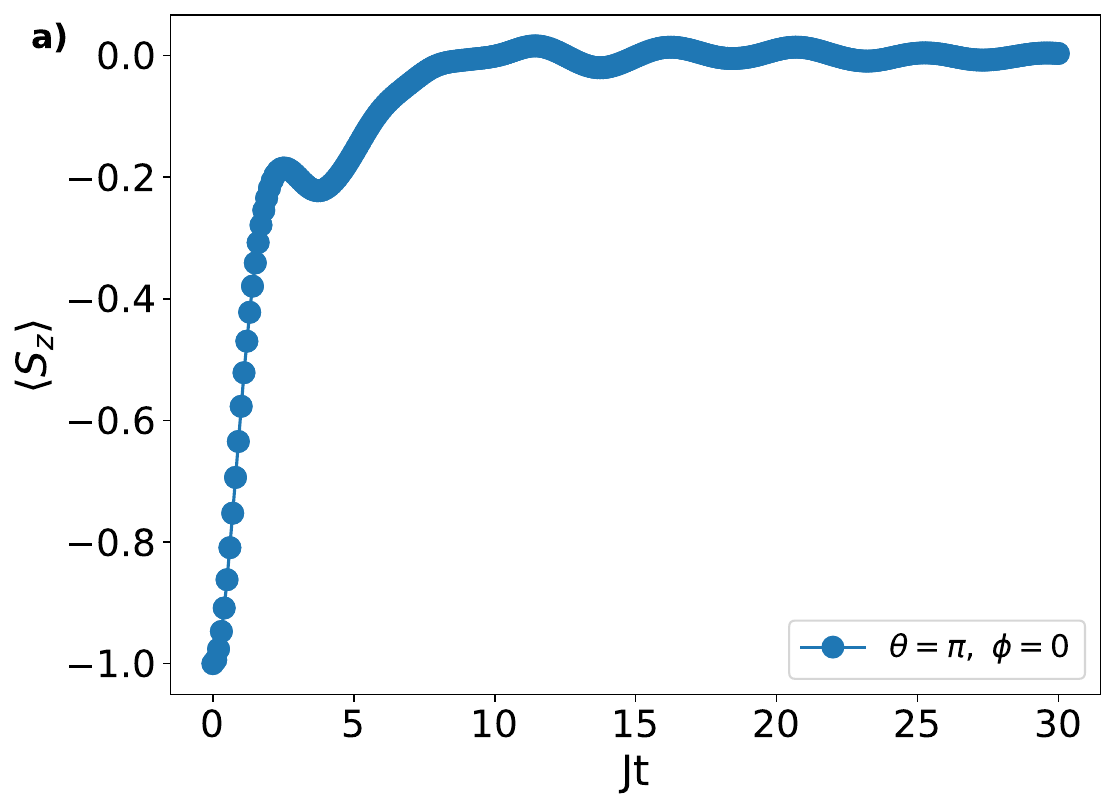}
\includegraphics[width=.45\textwidth]{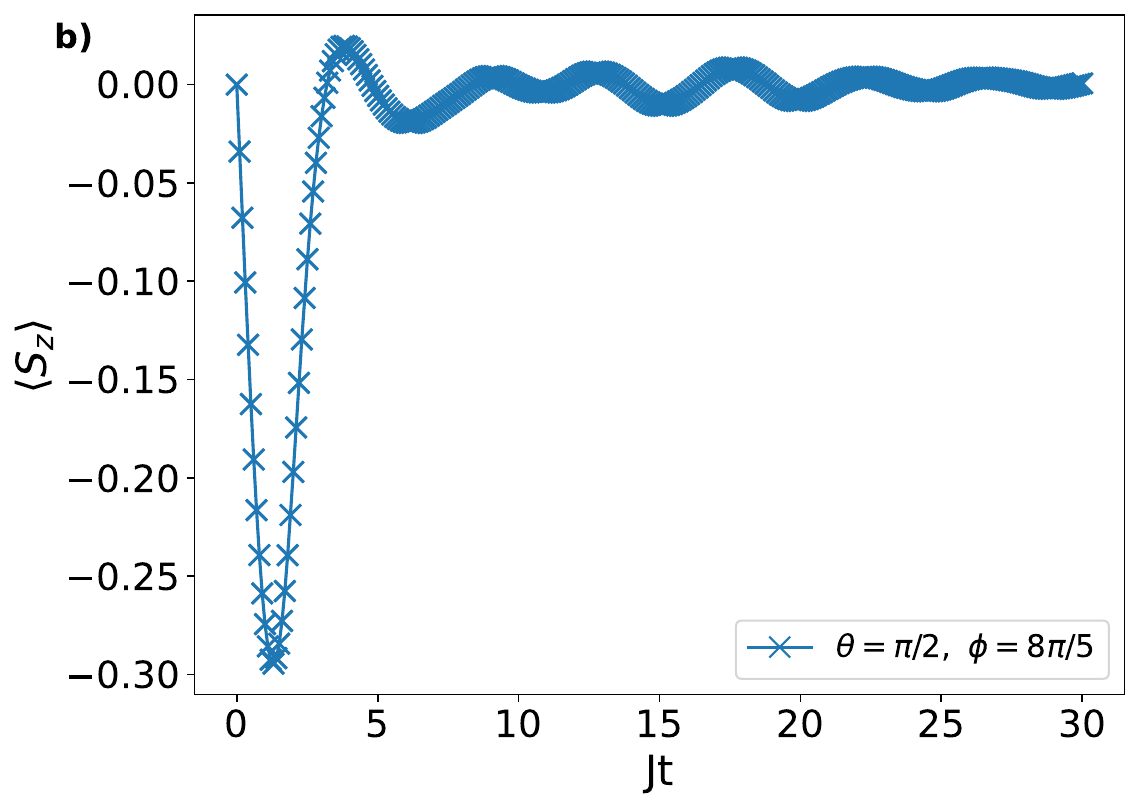}
\includegraphics[width=.45\textwidth]{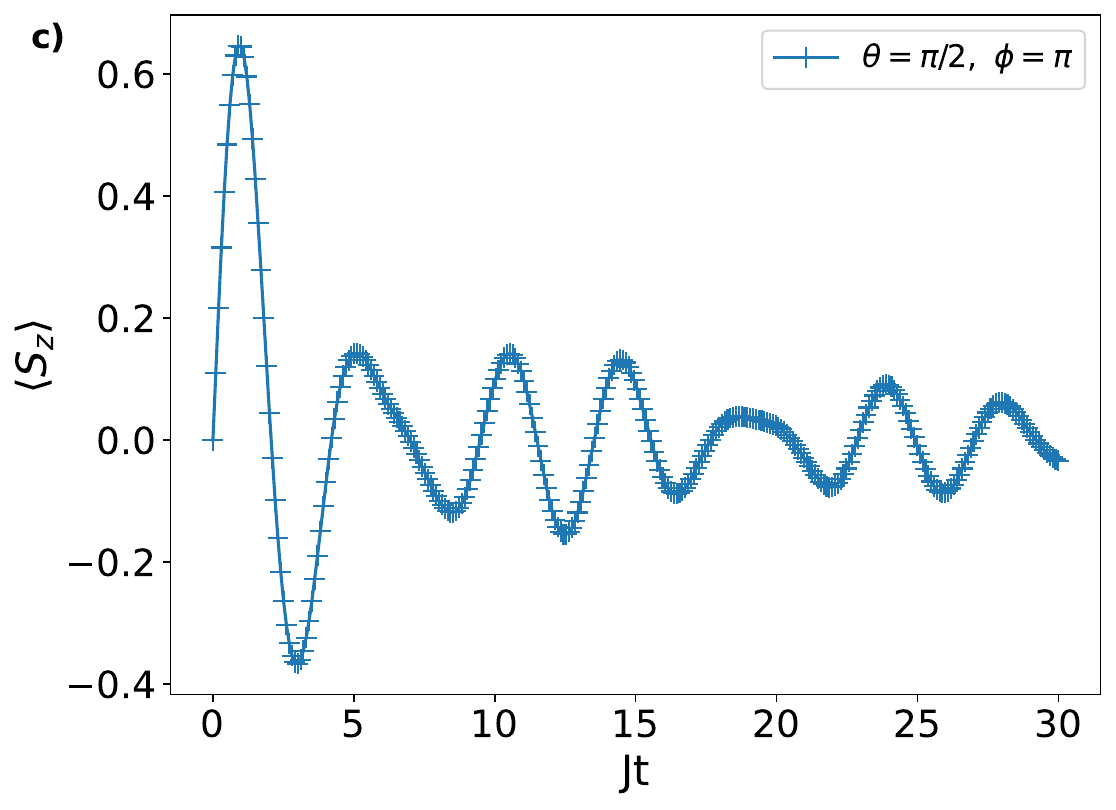}
\includegraphics[width=.45\textwidth]{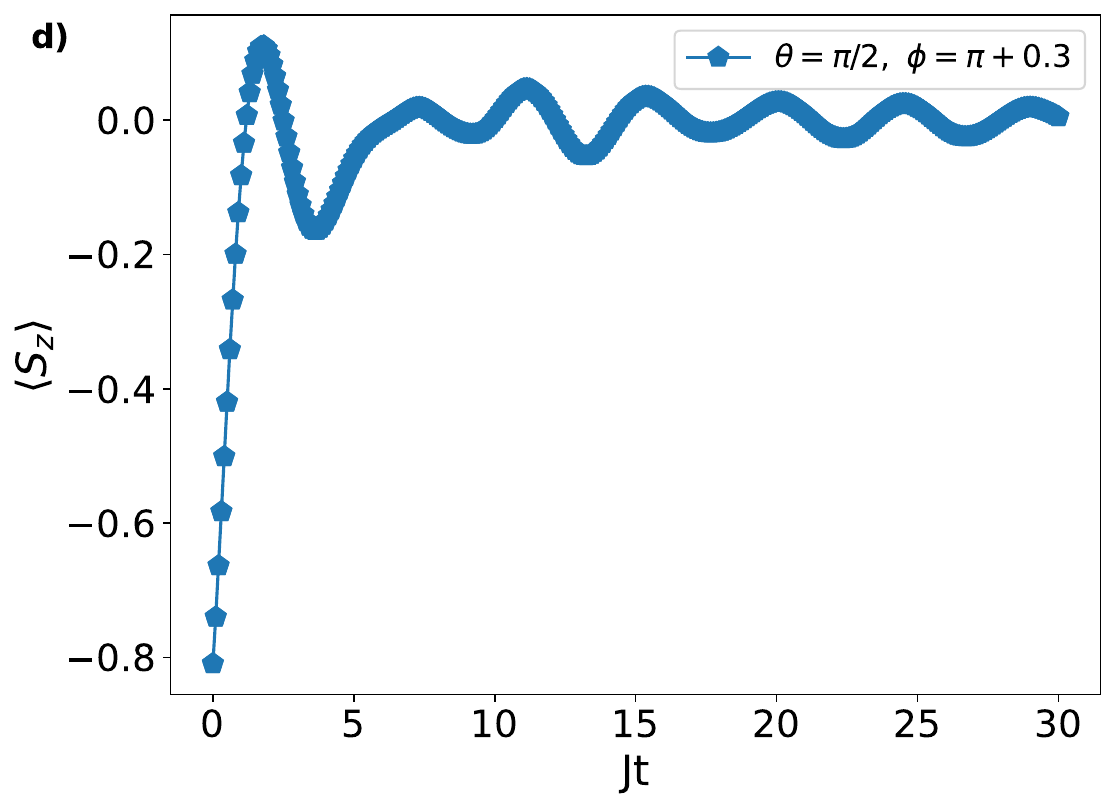}
\caption{Time evolution of the magnetization in the $z$ direction for
four different initial conditions showing the regimes of weak and strong thermalization. \textbf{(a and b)} Two initial states which strongly thermalize. They were studied in \cite{Fshen2021}. \textbf{c}: An initial state which weak thermalizes: the oscillations do not become small, but are around a thermal value. \textbf{(d)} an initial state with a $0.3$ increase in $\phi$ angle, showing that a small change in the initial state may change the regime of
thermalization}
\label{fig:mag}
\end{figure*}

In most works trying to connect equilibration and thermalization with a break of integrability and chaos, through ETH, $|\psi_0\>$ is the ground state of an initial Hamiltonian $H(\lambda_0)$.
The system Hamiltonian is then suddenly changed, quenched, to $H(\lambda)$ and we let the system
evolve isolated. Usually, one study under which $H$ and quenches the system thermalize
and try to connect with a break of integrability and quantum chaos. In these works the main question is if the equilibrium state is a thermal state; in general one directly looks if the system thermalizes or not. Consider the micro-canonical
thermal state
\begin{equation}
    \rho_{micro} =  \frac{1}{d_{E_0}} \sum_{E_n \in [E_0 \pm \Delta E]} |E_n\>\<E_n|
\end{equation}
with $d_{E_0}$ the Hilbert space dimension of the micro-canonical energy shell of size
$\Delta E$, such that it is small at the macroscopic scale, but still contains many
energy levels. One then studies
if $\overline{\rho}$ is close to $\rho_{micro}$ or if \footnote{In fact, von Neumann already analyzed the conditions for time averages to be equal to microcanonical in an attempt to obtain a quantum ergodic theorem \cite{Goldstein10}}
\begin{equation}
\<A\>_{micro} \approx \overline{\<A\>}.
\end{equation}

Recently, the main studied mechanism for thermalization is the eigenstate thermalization hypothesis (ETH),
where each individual energy eigenstate looks thermal:
\begin{equation}
\<E_n|A|E_n\> \approx \<A\>_{micro}, 
\end{equation}
for $E_n$ in an energy shell of size $\Delta E$. If ETH is valid, then the equilibrium value is the thermal one (when equilibration happens, see discussion below). Thus the question reduces to
which system obeys ETH. One scenario where ETH is valid, and where
analytical calculations are possible, is for random observables or $H$ \footnote{In these cases there is a connection between ETH and Typicality arguments, already made by von Neumann see Sec. 4.2.2 of \cite{Alessio16}
and \cite{Reimann15,Sugimoto23} for more recent works.}. In fact, for an $|E_n\>$ to obey ETH it has to be a "random" vector; its components on a given basis are randomly distributed. Apart from that, one has to consider exactly solvable models and numerical simulations of specific models, and there are many works \cite{Alessio16, Borgonovi16}.  Note that, in general, ETH is not valid for all $|E_n\>$, but for most of them apart from the borders of the energy spectrum; this is called weak ETH. 

For observables, ETH postulates the following anzats for the matrix elements of $A$ in the energy basis:
\begin{equation}\label{eq:eth}
    \hat A_{nm} = \delta_{n,m} A(\mathcal{E}) +  e^{-S\mathcal{(E})/2}f(\mathcal{E,\omega})B_{nm},
\end{equation}
where $A(\mathcal{E})$, $f(\mathcal{E,\omega)}$ are smooth functions of
average ($\mathcal{E}$) and difference ($\omega$) between energy $E_n$ and $E_m$, $S(\mathcal{E})$ is the thermodynamic entropy and $B_{nm}$ is a random real or complex variable with zero mean and unit variance. Thus one sees that to have ETH, in the energy basis, the diagonal elements of $A$ should not vary much and the off-diagonal should be uncorrelated and decay exponentially with the "distance" from the diagonal.

Thus, if ETH is valid on the whole spectrum the average state obtained from any initial state will be very close to the thermal micro-canonical state. This guarantees that any initial state will oscillate around the thermal state, but does not imply that the oscillations will be small most of the time \footnote{Note that the fluctuations depend on the off-diagonal part and equilibration will happen for any initial state
only if $\max B_{nm} \ll 1$; in this case if the $d_{eff}$ is not small the oscillations do not dephase, but their amplitude is undetectable.
In \cite{Reimann15} it is shown that most, or typical, observables will equilibrate and thermalize for any initial states. Weak thermalization is in this case an untypical situation.}.

\section{Weak and Strong Thermalization}

As mentioned before there are many studies about the connection between integrability and thermalization, and usually when the system is integrable there are local conserved quantities that prevent thermalization. In this case, the equilibrium state is a generalized thermal ensemble, that does not depend only on the average energy, but also
on other conserved quantities (see Sec. 8 of \cite{Alessio16}). There is also the case of pre-thermalization when the system first equilibrates to a non-thermal state, and only after a long time evolves to the thermal state
(see Sec. 8.4 of \cite{Alessio16}). 

More recently, using numerical simulations for the Ising model in a transverse and parallel field, another regime has been found and named weak thermalization \cite{Banuls2011}. 
In weak thermalization, while the average state is thermal, the initial large fluctuation does not seem to become small as happens in strong thermalization \footnote{In\cite{Banuls2011} is is claimed that strong thermalization can not happen classically, but note that even single classical states, may typically appear thermal, without any time average as proposed already by Boltzmann \cite{Lebowitz93}}. Latter, a mechanism to explain weak thermalization in terms of quasi-particle interactions was proposed in \cite{Chengju2017}.
It was also argued that the weak thermalization was due to the small number of quasi-particles, but that equilibration would eventually happen after longer times. Also, in \cite{Hyungwon15} it was numerically shown that there are local operators whose fluctuations decay slowly and that this could explain the weak thermalization of \cite{Banuls2011}. Finally, in \cite{Farrelly17} the authors obtain a lower bound on $d_{eff}$, for local $H$ and initial states with exponentially decaying correlation, that depends on the energy variance. It is then argued that the bound guarantees that equilibration will also happen in the weak thermalization regime after a long enough time.

Therefore, there is strong evidence that the phenomenon of weak thermalization is due to the time scales for the fluctuations to become small. However, after enough time are the fluctuations of weak and strong thermalization of the same magnitude? Most analyses up to date were more qualitative. And it is important to have a more quantitative analysis, in particular studying how the size of the typical fluctuations decays with the system size; is there any particular behavior in the regime of weak thermalization? We will proceed to more quantitative analyses, studying the scaling of the size of fluctuations using $d_{eff}$ and show that in both, weak and strong thermalization, the fluctuations decay exponentially with $N$ but at different rates.

We should also mention that in \cite{Sun21} a connection between how much information spreads in the systems (scrambling) and weak-strong thermalization was studied using tripartite mutual information: weak thermalization was related to slow scrambling. It was also shown, that weak thermalization happens for initial states with energy near the edge of the spectrum. Interesting, it was also shown that weak thermalization also occurs in the XX model with parallel field and that this could be observed experimentally with superconducting qubits; the experiment was done in \cite{Fshen2021}. Weak thermalization was also found in a long-range Ising chain \cite{Liu19}.

\section{Results}
\label{sec:results}

\begin{figure}
\includegraphics[width=.51\textwidth]{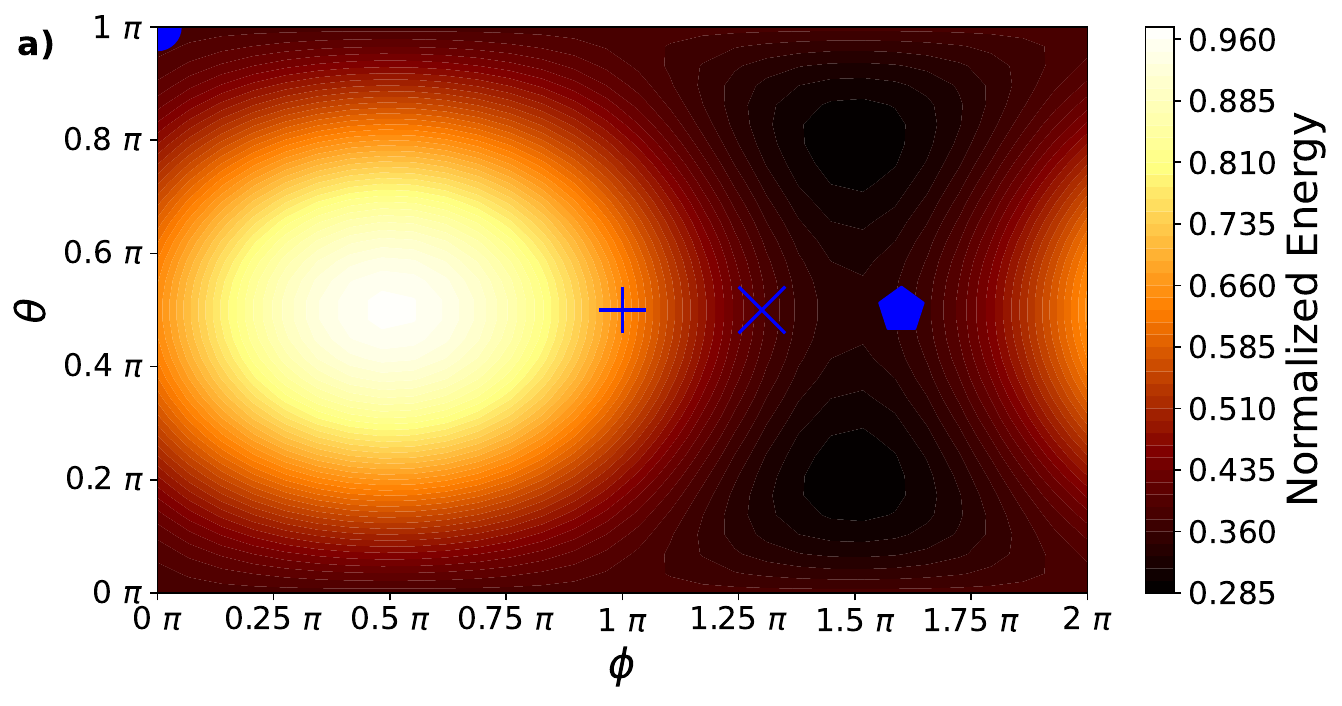}
\includegraphics[width=.51\textwidth]{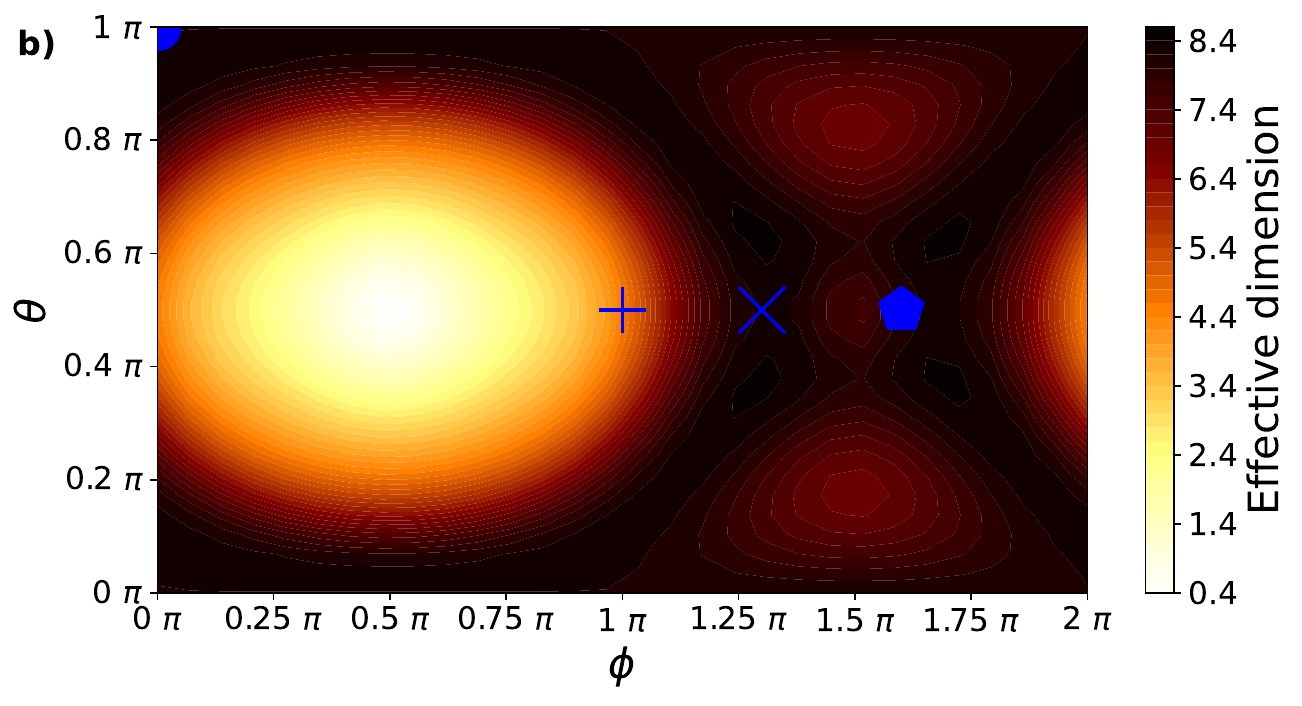}
\caption{\textbf{Upper panel:} Normalized Energy (NE) as a function of $\theta$ and $\phi$, for initial states with $N =15$. \textbf{Bottom panel:} Logarithm of the effective dimension as a function of $\theta$ and $\phi$, for $N =15$. 
One can see a complete equivalence between large effective dimension and small NE (dark regions), both indicating strong thermalization. The markers on the plot indicate the initial states depicted in Fig.\ref{fig:mag} with the same markers.}
\label{fig:effdim}
\end{figure}

\begin{figure}
\centering
\includegraphics[width=.49\textwidth]{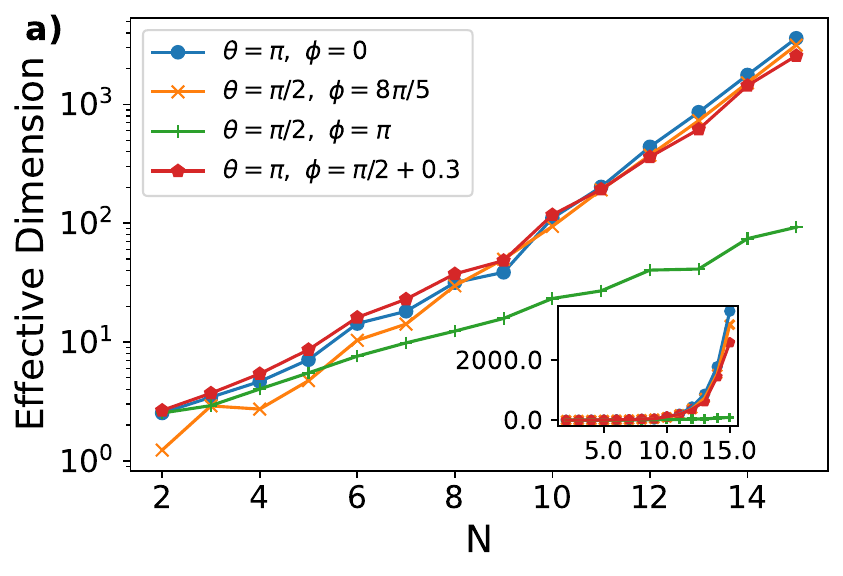}
  \includegraphics[width =.49 \textwidth]{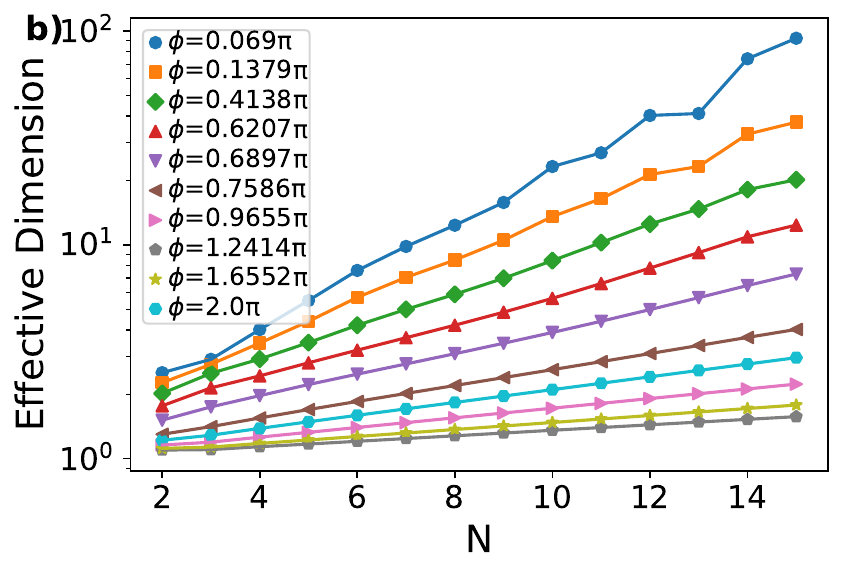}
\includegraphics[width=.49\textwidth]{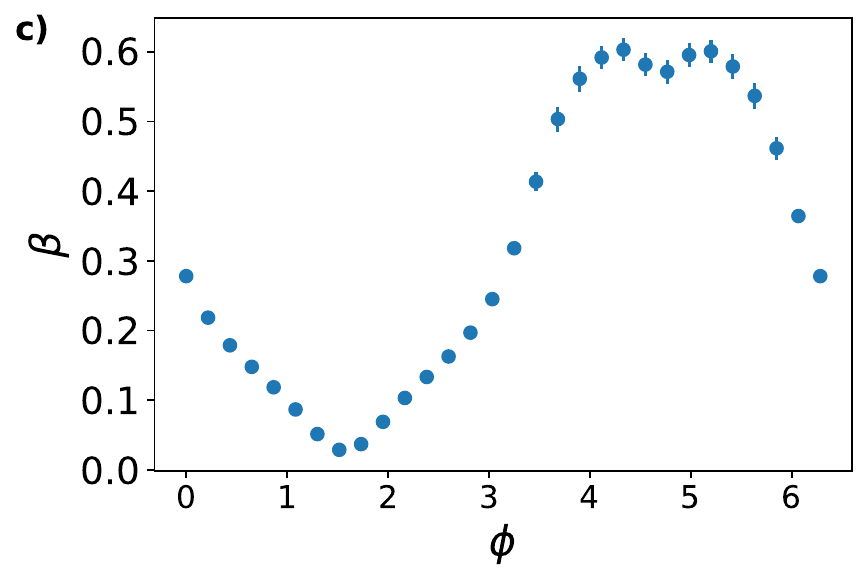}
\caption{\textbf{Upper panel}: Effective dimension plotted as a function of the number of spins in a semi-log scale (inset shows in a regular scale) for the same four initial conditions depicted in Fig.\ref{fig:mag}. We can see
that the effective dimension grows faster for the states that equilibrate. \textbf{Middle Panel}: Each line represents the growth of effective dimension for different values of $\phi$ for a fixed $\theta = \pi/2$ also in a semi-log scale (inset shows regular scale),  note that the slope is not a monotonic function of $\phi$. For instance, the small ones are in the brighter regions in Fig.\ref{fig:effdim}. \textbf{Bottom panel}: The exponent of effective dimension growth with the number of spins calculated as a function of $\phi$ for the lines of Fig.\ref{fig:eff} (b) and other values; error bars were calculated from the covariant matrix.
It seems the transition between weak and strong thermalization has no discontinuity.}
\label{fig:eff}
\end{figure}

\begin{figure}
    \centering
    \includegraphics[width=.49\textwidth]{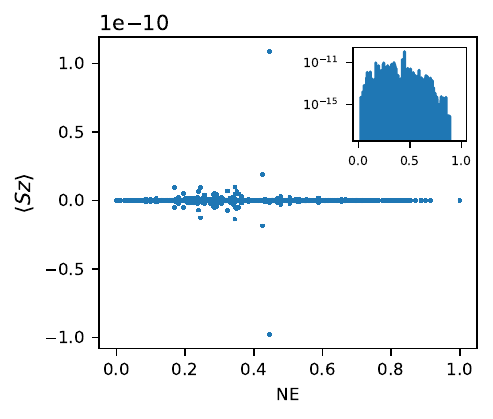}
    \caption{Expectation values of each eigenstate of the Hamiltonian, Eq.\eqref{eq:prlH} (for $N=12$) as a function of the associated normalized energy (NE). Each value below $10^{-10}$ can be considered as null. In this plot, insets depict the plot in a semi-log scale.}
    \label{fig:eignmag}
\end{figure}

We will consider the XX model in a parallel field, that is non-integrable, was studied numerically in \cite{Sun21}, experimentally in \cite{Fshen2021}, and is given by
\begin{equation}\label{eq:prlH}
    H = \frac{J}{2} \sum_{j = 1}^{N} (\sigma_j^x\sigma_{j+1}^x + \sigma_j^y\sigma_{j+1}^y) + g  \sum_{j = 1}^{N} \sigma_j^y.
\end{equation}
The initial conditions are product spin states given by 
\begin{equation}\label{eq:isostates}
    \ket{\theta, \phi} = \prod_{j=1}^{N} \text{cos} \frac{\theta}{2} \ket{Z_+}_j + e^{-i\phi}\text{sin} \frac{\theta}{2} \ket{Z_-}_j, 
\end{equation}
where $\ket{Z_\pm}_j$ is the eigenbasis of the $\sigma_j^z$ Pauli matrix. Here we use $J = 1$ and $g = J/2 + 0.01.$ All our simulations are done with exact diagonalization for chains up to $N= 15$ spins.

We will first reproduce some of the results of \cite{Fshen2021},
to latter compare with ours. To qualitatively illustrate the difference between weak and strong thermalization, as done in \cite{Fshen2021},
we show in Fig.\ref{fig:mag} the time evolution of the expectation value the total magnetization in $z$ direction, for different initial states defined in \eqref{eq:isostates}. We can see in Figs.\ref{fig:mag}(a) and \ref{fig:mag}(b) that for some initial states, after a fast transient time, the observable oscillations amplitude becomes small; the system strongly thermalizes. But for other initial states, Fig.\ref{fig:mag}(c), the fluctuations keep oscillating with significant amplitude. Note also that a small change in the initial state may change the behavior Fig.\ref{fig:mag}(d).

As suggested in the supplementary material of \cite{Banuls2011}, for another model, and studied in \cite{Fshen2021} for this model, this behavior is related to the behavior of entanglement between one spin and the rest of the chain (how it increases with time and its time average value) and to the position of the average energy of the initial state in relation to the energy spectrum of $H$. 

To quantify the position of the energy of the initial state in relation to the spectrum of $H$ one can use the Normalized Energy (NE):
\begin{equation}
    \text{NE}(\theta,\phi) = \frac{\bra{\theta,\phi} H \ket{\theta,\phi} - \text{min} (E_\alpha)}{\text{max}(gap)},
\end{equation}
where max$(gap) = \text{max}(E_\alpha) - \text{min}(E_\alpha)$. In Fig.\ref{fig:effdim}(a) we plot NE as a function of $\theta$ and $\phi$, reproducing the results
of \cite{Sun21,Fshen2021}. It can be seen that weak thermalization happens for initial states with energy close to the border of the spectrum (white and yellow regions), while strong thermalization happens in the middle of the spectrum (black and reddish regions).

We now present our first result, showing that the $d_{eff}$ indicates if
an initial state weak or strongly thermalizes. This can be seen in Fig.\ref{fig:effdim}(b), where we plot the logarithm of $d_{eff}$ instead of NE \footnote{We choose to plot the logarithm of the effective dimension to have a better visualization.}. One can see that $d_{eff}$ also identifies the regimes of weak and strong thermalization: weak thermalization happens for small $d_{eff}$ (white and yellow regions) and strong for large $d_{eff}$ (black and reddish regions). Note that weak thermalization occurs in the same regions on both graphs. The markers in the plot identify the initial conditions studied in Fig.\ref{fig:mag}. We can see there that by fixing $\theta = \pi/2$ we can have different regimes of thermalization as a function of $\phi.$ The initial states of all spins pointing up are depicted in Fig.\ref{fig:mag}(a), and the remaining initial states are in the line $\theta = \pi/2$.
Naturally, $d_{eff}$ decreases as the energy of the initial state goes to the border of the spectrum, since the variance of energy, and therefore $d_{eff}$ has to decrease in this limit; if the initial state energy is the maximum eigenstate the variance is null.  It is also known that ETH usually breaks at the border of the spectrum, but it is unclear whether these two phenomena are related. In our case, as shown below, ETH is valid on the border, while equilibration ceases. We should mention that in the appendix D of \cite{Sun21} the distribution of the coefficients $c_{n}$ was studied for a fixed system size and different initial states; while no connection was made with the $d_{eff}$ its behavior is related to these distributions. Note that, since $d_{eff}$ is an upper bound on the fluctuations, it is a sufficient but not necessary, indicator of equilibration.

We also briefly study the ETH by looking at the diagonal elements of magnetization on the energy basis. As we can see in Fig.\ref{fig:eignmag}, the value of magnetization in $z$ direction for all eigenstates of $H$ was numerically calculated and they are of order $\mathcal{O}(e-10)$. It perfectly fits the case when ETH is valid over the whole spectrum, i.e. the value of the observable is thermal for all eigenstates, and justifies why the time averages of magnetization are equal to the thermal value for any initial state. Note that if ETH is not valid
at the border of the spectrum, one could have an initial state that equilibrates but does not thermalize.

As mentioned before, to study more quantitatively if a system equilibrates or thermalizes it is important to analyze systems of different sizes since we do not expect such behavior for small systems. In particular, it is important to analyze how the size of the typical fluctuation, after a long long time, scales with $N$. Thus, we study how $d_{eff}$, an upper bound on the size of the typical fluctuation, increases with $D=2^N$, as shown in Fig.\ref{fig:eff}(a) for the four initial states of Fig.\ref{fig:mag}. It can be seen, that for weak and strong thermalization $d_{eff}$ increases exponentially with $N$.
In fact, this is the behavior expected for random initial states and $H$, while integrable systems have a polynomial behavior \cite{Venuti13}. Therefore, the difference between weak and strong thermalization is in the coefficient of the exponential: in strong thermalization $d_{eff}$, and therefore the fluctuations, decay faster with $N$ than in weak thermalization. 

One interesting question is whether there is a sharp transition between weak and strong thermalization regimes. To study this,
we analyzed the scaling of $d_{eff}$ for $\theta = \pi/2$ and different values of $\phi$ in Fig.\ref{fig:eff}(b) in a semi-log scale. One can see that all the curves are seen to obey an exponential increase, as seen for the four initial states of Fig.\ref{fig:mag}. 
To have a more quantitative analysis, we obtain the exponent $\beta$ from a fit of $d_{eff}\sim e^{\beta N}$ for
each curve of Fig.\ref{fig:eff}(b). This is shown in Fig.\ref{fig:eff}(c), where a non-monotonic behavior can be seen (as expected from Fig.\ref{fig:effdim}) and no sharp transition
as we change $\phi$. Therefore it seems there is no clear cut-off between weak and strong thermalization, but a smooth crossover.
Note that, in principle for any $\beta >0$ the system will equilibrate for large enough $N$. Therefore in principle, the difference between strong and weak thermalization is only on the size scales needed to obtain thermal behavior;
and also on the time scales as argued in \cite{Chengju2017,Hyungwon15,Farrelly17}. Note that the 
$\beta$ can be close to zero, in the brighter part of Fig.\ref{fig:effdim}, but it is not zero as the initial state would be an eigenstate of $H$.

\section{Conclusion}

In the last decades, the study of equilibration and thermalization of isolated quantum systems attracted a lot of attention, leading to at least two parallel and
complementary lines of research: one more abstract with rigorous theorems about generic systems and another studying many particular but physically motivated models. In particular, in a numerical study it was shown that for a non-integrable spin chain model there are initial states with fluctuations around the thermal state that seem to persist over time; a phenomenon named weak thermalization. Later, it was argued that the fluctuations would decay eventually after a long time \cite{Chengju2017, Hyungwon15, Farrelly17}. However, there was no quantitative study of the sizes of the fluctuations in weak and strong thermalization, but only a proposal that weak thermalization was due to the behavior of quasi-particles \cite{Chengju2017} and the average energy of the initial state \cite{Sun21,Fshen2021}. Here we analyzed this phenomenon using the scaling of effective dimension, a quantity that appears in rigorous theorems about the equilibration of isolated quantum systems. We showed that the effective dimension is a simple way to explain weak and strong thermalization.
Furthermore, we gave evidence that the size of the fluctuations decays exponentially with system size in both regimes, indicating a smooth crossover between the two regimes and that thermalization will happen for large enough systems in both cases. Finally, we pointed out that ETH being valid in the whole spectrum allows for the thermalization of any initial state, although it may be weak or strong.

Finally, we have shown that the rigorous mathematical bounds on the size of the fluctuations \cite{Reimann2008, Linden08, Gogolin16} can be useful to understand weak and strong thermalization, in particular through the effective dimension. We hope this work helps to establish more connections between these parallel lines of research.
In fact, the effective dimension is already used in much
condensed matter works under the name of inverse participation ratio (IPR), in particular in the study
of many-body localization, without relating it to the rigorous mathematical bounds (see for example, \cite{Tobias19}). Also, we believe that the quasi-particle explanation is related to the dephasing picture of equilibration and a connection can lead to a better estimation of the equilibration time \cite{Tro2018}. It is also important to use other numerical methods to approximate larger systems and study the crossover in such cases.

\begin{acknowledgments}
		
\end{acknowledgments}
We thank Fernando Iemini for valuable discussions and insights about the problem. This work is supported by the Instituto Nacional de Ciência e Tecnologia de Informação Quântica ((465469/2014-0), and by the Air Force Office of Scientific Research under award number FA9550-23-1-0092.

\bibliography{Equilibration-WeakThermalization}

\appendix

\end{document}